\documentclass[]{spie}

\usepackage{amsmath}
\usepackage{graphicx}
\usepackage{booktabs}
\usepackage{siunitx}
\usepackage{url}
\usepackage{subcaption}
\usepackage{placeins}

\title{Simulation-based dynamic pointing analysis of AtLAST under wind loading and fast-scan conditions}

\author[a]{Aleksej Kiselev}
\author[a]{Martin Timpe}
\author[a]{Matthias Reichert}
\author[b,c]{Tony Mroczkowski}
\author[d]{Claudia Cicone}

\affil[a]{OHB Digital Connect, Weberstra\ss e 21, D-55130 Mainz, Germany}
\affil[b]{Institute of Space Sciences (ICE-CSIC), Carrer de Can Magrans, s/n, 08193 Cerdanyola del Vallès, Barcelona, Spain}
\affil[c]{Institut d'Estudis Espacials de Catalunya (IEEC), E-08860 Castelldefels, Barcelona, Spain}
\affil[d]{Institute of Theoretical Astrophysics, University of Oslo, PO Box 1029, Blindern, 0315 Oslo, Norway}

\authorinfo{Send correspondence to: aleksej.kiselev@ohb.de
}

\begin{document}

\maketitle

\begin{abstract}
The Atacama Large Aperture Submillimeter Telescope (AtLAST) is a next-generation 50-m class single-dish facility concept designed for high-throughput, wide-field mapping at millimetre and submillimetre wavelengths. Its science goals require fast telescope motion while maintaining stringent dynamic pointing stability under realistic environmental and operational conditions. We present an end-to-end dynamic pointing analysis of the current AtLAST design, coupling a flexible structural model of the telescope mount with the intended cascaded main-axis control architecture, pointing-error evaluation, and measured site wind excitation.

A key feature of the study is the use of high-rate wind measurements obtained at the AtLAST candidate sites before telescope construction. The measured wind time series are used as dynamic input to evaluate tracking under wind loading for multiple elevation angles, wind-speed classes, and wind angles of attack. The resulting wind-induced dynamic pointing error remains within the allocated tracking-stability budget for all budgeted wind conditions, confirming the adequacy of the current structural concept and main-axis control approach with respect to wind-driven pointing stability.

Fast mapping is investigated separately using a Lissajous-daisy scan close to the kinematic limits of the mount. In contrast to the wind-loaded tracking case, the scan-induced response is more strongly linked to finite control bandwidth and trajectory-following accuracy, particularly in elevation. The analysis therefore identifies two distinct dynamic pointing regimes and provides guidance for future control optimization, feedforward strategies, and active compensation concepts.
\end{abstract}


\section{Introduction}

The Atacama Large Aperture Submillimeter Telescope (AtLAST) is being developed as a next-generation single-dish facility for wide-field observations at millimetre and submillimetre wavelengths. The telescope concept combines a 50-m aperture with a degree-scale field of view, high mapping speed, and operation across the full ground-accessible submillimetre range. These capabilities are required to address science cases that rely on large-area, high-throughput mapping, including surveys of dust-obscured galaxy evolution, the circumgalactic and intergalactic medium, Galactic star formation, Solar physics, planetary science, and time-domain astronomy~\cite{mroczkowski2025conceptual,cicone2026overview}.

The same science requirements that motivate AtLAST also impose demanding constraints on the telescope dynamics. Efficient wide-field mapping requires fast and repeatable telescope motion, with the current design targeting scan speeds up to $3\,\mathrm{deg\,s^{-1}}$ and accelerations up to $1\,\mathrm{deg\,s^{-2}}$~\cite{cicone2026overview}. At the same time, observations at the shortest wavelengths require tight control of the optical performance, including surface accuracy, alignment, and pointing stability. For a telescope with a 50-m reflector, a large secondary mirror, an exposed structure, and a total mass on the order of several thousand tonnes, small dynamic deformations can translate into measurable pointing errors. Dynamic pointing performance is therefore a system-level design driver rather than a secondary operational effect.

Dynamic pointing errors arise from the coupled response of the flexible telescope structure, the main-axis control system, and the environmental or operational excitation. During tracking, wind loading produces time-dependent structural deformation and residual pointing jitter. During fast mapping, the commanded scan motion itself becomes an excitation source: the main-axis controller generates actuator torques to follow the trajectory, and these torques can excite flexible structural modes. Similar end-to-end dynamic simulation and jitter-budget approaches are used for other next-generation large telescope mounts, where wind-induced jitter and self-induced vibrations are treated as key contributors to tracking performance~\cite{ippa2024gmt_jitter}. Static structural analysis or quasi-static wind-load assessment alone cannot capture these coupled effects. A time-domain end-to-end simulation is required to evaluate the closed-loop telescope response at the optical pointing level.

This paper presents such an end-to-end dynamic pointing analysis for the current AtLAST design. The simulation framework couples a flexible state-space representation of the telescope mount with the cascaded main-axis control architecture foreseen for AtLAST and evaluates the resulting pointing error in elevation and cross-elevation. Two operating regimes are analysed separately. First, tracking under wind loading is simulated using measured wind time series as external dynamic loads. Second, fast-scan operation is represented by a Lissajous-daisy trajectory close to the kinematic limits of the mount. This separation allows the wind-driven and self-induced contributions to dynamic pointing error to be interpreted independently.

A distinctive aspect of the analysis is the use of high-rate wind measurements obtained at the AtLAST candidate sites before telescope construction. The AtLAST site-characterization campaign includes meteorological towers equipped with sonic anemometers, providing time-resolved wind data at sampling rates up to $20\,\mathrm{Hz}$~\cite{atlast_site_selection_report,pizarro_inprep,cicone2026overview}. Such site-specific data are rarely available at this stage of a large-telescope design and provide a direct link between site characterization and dynamic pointing verification.

The objective of this work is therefore twofold. First, the analysis verifies whether the current AtLAST structural concept, in combination with the intended main-axis control architecture and realistic site wind excitation, satisfies the allocated wind-induced dynamic pointing budget. Second, it identifies the dominant mechanisms that drive self-induced pointing errors during fast scanning. The results show that the wind-loaded tracking case remains within the allocated budget and is mainly governed by residual flexible structural response, whereas the fast-scan case exhibits a stronger dependence on finite control bandwidth and trajectory-following accuracy, particularly in elevation. The resulting decomposition provides design-relevant guidance for future control optimization, feedforward strategies, and active compensation concepts.

\section{End-to-end dynamic pointing simulation framework}
\label{sec:e2e_framework}

The dynamic pointing performance of a large, flexible telescope cannot be assessed from the structural model, the control system, or the disturbance environment in isolation. Wind-induced deformation, finite control bandwidth, structural resonances, and scan-induced inertial forces interact dynamically and jointly determine the pointing error. For this reason, the AtLAST dynamic pointing performance is evaluated using an end-to-end simulation framework that couples the flexible telescope mount, the main-axis control system, realistic wind excitation, commanded scan trajectories, and the pointing evaluation in the time domain.

The purpose of the framework is to predict the dynamic pointing error at the telescope boresight under representative operational conditions. The pointing error is evaluated as the deviation of the telescope line of sight and is expressed in elevation and cross-elevation coordinates. This distinction is important because the controlled main-axis coordinates and the pointing coordinates are not identical in a flexible structure. Elastic deformation between the drive encoders and the optical boresight can generate residual pointing error even when the axis tracking error remains small.

\subsection{Simulation concept}
\label{subsec:simulation_concept}

The end-to-end model consists of three coupled elements: the flexible structural model of the telescope mount, the main-axis control system, and the pointing evaluation. The structural model describes the dynamic response of the mount to actuator torques and external loads. The control system generates azimuth and elevation drive torques from the commanded and measured axis positions. The pointing evaluation maps the resulting rigid-body motion and elastic deformation to pointing errors in elevation and cross-elevation.

Within the simulation, the flexible mount acts as the dynamic plant of the azimuth and elevation control loops. The controller receives commanded axis trajectories and feedback signals from the plant and returns actuator torque commands. These torques are applied to the flexible structural model, together with external loads where applicable. The same framework can therefore be used for tracking simulations under wind loading and for fast scanning simulations without changing the underlying plant or control model.

A central feature of the framework is that the pointing error is evaluated at the optical boresight rather than solely at the controlled axes. This allows the total dynamic pointing error to be separated into a control-related contribution, associated with finite tracking accuracy of the main axes, and a residual contribution, dominated by flexible structural deformation.

\subsection{State-space representation of the flexible mount}
\label{subsec:state_space_mount}

The flexible telescope mount is represented by a linear time-domain model derived from the finite-element modal representation of the current AtLAST structural design. The model retains the dynamic behaviour relevant for the pointing analysis and includes both the rigid-body motion of the main axes and the flexible structural modes that can be excited by actuator torques or external wind loads.

For integration into the end-to-end simulation, the modal structural model is transformed into a first-order state-space representation,

\begin{equation}
\dot{\mathbf{x}}(t)
=
\mathbf{A}\mathbf{x}(t)
+
\mathbf{B}_{u}\mathbf{u}(t)
+
\mathbf{B}_{w}\mathbf{w}(t),
\label{eq:state_space_model}
\end{equation}

\begin{equation}
\mathbf{y}(t)
=
\mathbf{C}\mathbf{x}(t)
+
\mathbf{D}_{u}\mathbf{u}(t)
+
\mathbf{D}_{w}\mathbf{w}(t),
\label{eq:output_equation}
\end{equation}

where $\mathbf{x}$ contains the modal coordinates and modal velocities, $\mathbf{u}$ denotes the actuator torque inputs from the main-axis drive system, and $\mathbf{w}$ contains external disturbance loads, in particular the time-dependent wind loads. The output vector $\mathbf{y}$ contains the quantities required for closed-loop simulation and performance evaluation. These include the simulated encoder signals used by the azimuth and elevation controllers, as well as pointing deviations expressed in elevation and cross-elevation.

This formulation is central to the end-to-end analysis because the controlled mechanical coordinates and the pointing coordinates are not identical in a flexible telescope. The controller acts on the main-axis feedback signals, while the pointing performance is evaluated at the optical boresight. Flexible deformation between these locations can therefore generate residual pointing error even when the controlled axis positions follow their references accurately.

\subsection{Simulation cases}
\label{subsec:simulation_cases}

Two simulation cases are considered separately in order to distinguish the dominant physical mechanisms contributing to the dynamic pointing error. The first case addresses wind-induced pointing jitter during tracking. In this case, the commanded azimuth and elevation positions are kept constant by the control system, while measured wind time series are applied as external dynamic loads. The resulting pointing error quantifies the telescope response to realistic atmospheric excitation under closed-loop tracking conditions.

The second case addresses self-induced pointing jitter during fast scanning. In this case, no wind load is applied. Instead, time-dependent azimuth and elevation trajectories are prescribed. The self-induced jitter is therefore not introduced as an independent disturbance input, but arises inherently from the closed-loop response to the commanded scan motion. The actuator torques generated by the control system excite structural modes of the flexible mount and lead to dynamic pointing errors in elevation and cross-elevation.

The separation of wind-induced and scan-induced effects is a deliberate modelling choice. During real observations, fast scanning may occur simultaneously with wind loading. However, analysing both effects independently allows their relative magnitudes and physical origins to be assessed separately. This provides a basis for identifying whether dynamic pointing is primarily limited by external wind excitation, by finite control bandwidth, by scan-driven excitation of structural modes, or by a combination of these effects.

\subsection{Pointing-error definitions}
\label{subsec:pointing_error_definitions}

The pointing error is evaluated in elevation and cross-elevation coordinates. For each coordinate, the total pointing-error time series is separated into a static and a dynamic component. The static pointing error is defined as the mean value of the pointing-error signal,

\begin{equation}
PE_{\mathrm{static}}
=
\overline{PE(t)} ,
\label{eq:static_pointing_error}
\end{equation}

whereas the dynamic pointing error is defined as the root-mean-square deviation from this mean value,

\begin{equation}
PE_{\mathrm{dyn}}
=
\sqrt{
\frac{1}{T}
\int_0^T
\left(
PE(t) - \overline{PE(t)}
\right)^2
\,dt
}.
\label{eq:dynamic_pointing_error}
\end{equation}

The static component can in principle be compensated by a pointing model or calibration procedure. The dynamic component represents residual jitter during the observation and is therefore the relevant quantity for assessing tracking stability and scan performance.

In addition to the total pointing error, the simulation evaluates the position-control error of the main axes. This quantity describes the tracking error associated with finite control bandwidth. The remaining contribution is referred to as the residual pointing error and is dominated by flexible structural deformation. The decomposition is performed on the time-domain signals before RMS evaluation. Consequently, the RMS value of the residual component is not generally equal to the scalar difference between the RMS values of the total pointing error and the control error.

The position-control error is obtained from the commanded and measured main-axis motion and is therefore an observable quantity of the closed-loop system. This distinguishes it from the residual pointing contribution, which is associated with flexible structural deformation and cannot be inferred from the main-axis tracking error alone.

\section{Main-axis control design and tuning}
\label{sec:control_design}

The main-axis control system determines how the telescope mount rejects external disturbances and follows commanded trajectories. In the end-to-end framework, the azimuth and elevation axes are controlled independently by cascaded feedback loops acting on the flexible state-space model of the mount. The control model captures the dominant closed-loop dynamics relevant for dynamic pointing, including position tracking, disturbance rejection, actuator torque generation, and the interaction with flexible structural modes.

\subsection{Control architecture}
\label{subsec:control_architecture}

Both the azimuth and elevation axes are controlled by the cascaded main-axis control architecture foreseen for AtLAST. In the end-to-end simulations, this control architecture is coupled to the flexible state-space model of the current telescope design, forming a closed-loop telescope model for assessing dynamic pointing performance under representative operating conditions.

The control architecture consists of an outer position loop, an inner velocity loop, and a drive-torque stage. The position loop compares the commanded axis position with the simulated encoder position and generates a velocity demand. The velocity loop converts this velocity demand into a torque command, which is then passed through the drive-torque dynamics and applied to the flexible mount model through the actuator input matrix of the state-space representation.

\begin{figure}[ht]
    \centering
    \includegraphics[width=1\linewidth]{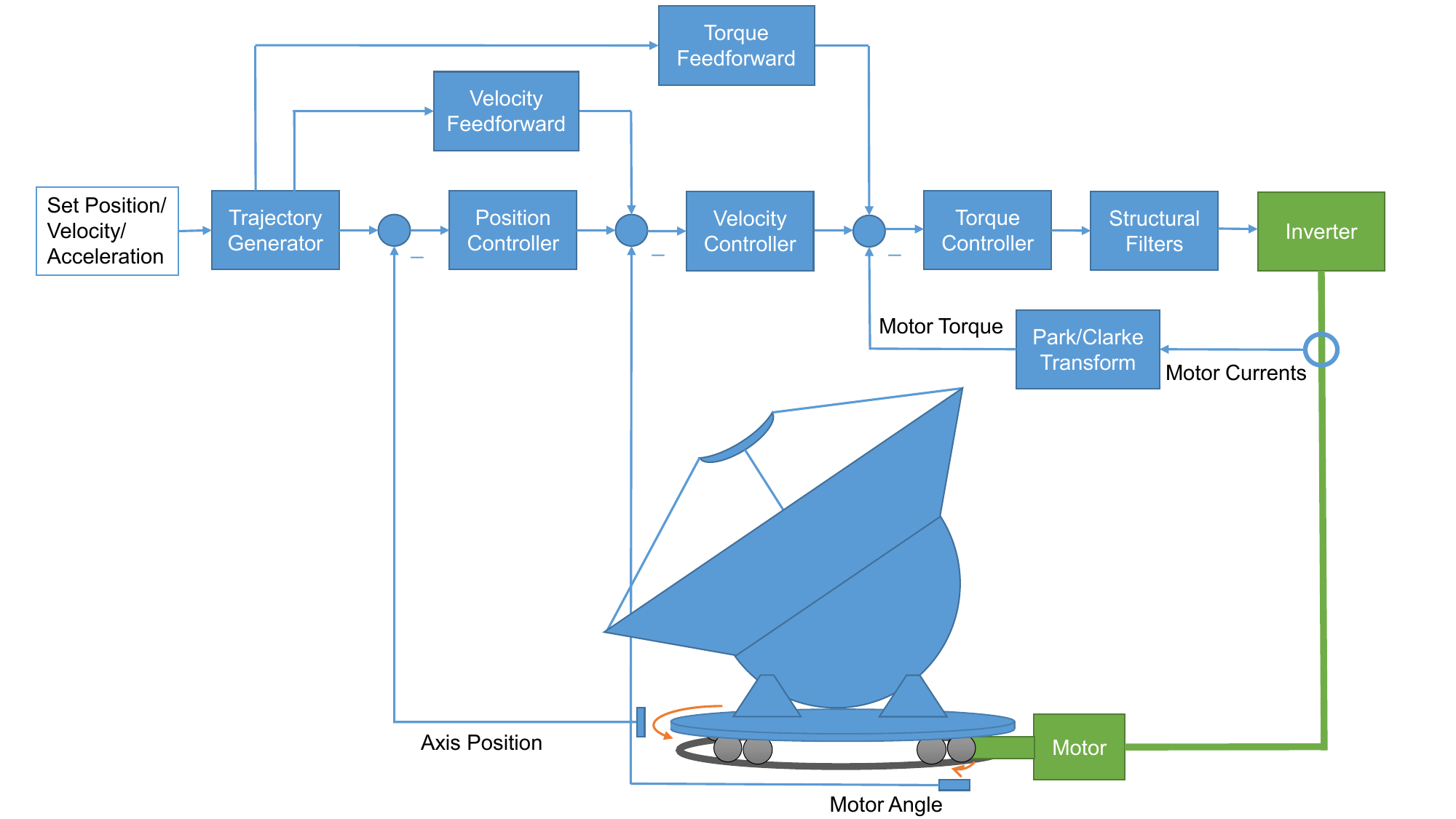}
    \vspace{.2cm}
    \caption{Cascaded main-axis control structure used in the end-to-end dynamic pointing simulations. The outer position loop generates a velocity demand from the axis position error, while the inner velocity loop and drive-torque dynamics generate the actuator torque applied to the flexible mount model.}
    \label{fig:main_axis_control_structure}
\end{figure}

A schematic representation of the implemented control structure is shown in Fig.~\ref{fig:main_axis_control_structure}. The azimuth and elevation axes use the same cascaded control concept, with axis-specific controller parameters. The finite bandwidth of these loops determines how effectively the telescope can reject disturbances and follow commanded motion, while the generated actuator torques provide the dynamic input through which the control system interacts with the flexible structural modes of the mount.

The drive-torque dynamics are represented by a first-order approximation including the relevant delay. This reduced representation captures the frequency range relevant for the dynamic pointing analysis while allowing efficient time-domain simulations with the flexible structural model. Feedforward terms can be included to improve trajectory tracking, while the primary stability and disturbance-rejection properties are determined by the cascaded feedback loops.

\subsection{Open-loop tuning approach}
\label{subsec:open_loop_tuning}

The main-axis controllers are tuned using open-loop frequency-response analysis of the controlled flexible mount model. For each axis, the loop is opened at the position-control level and evaluated for representative telescope elevations. Gain margin, phase margin, and position-loop bandwidth are extracted as robustness and performance metrics of the coupled telescope-control model.

The stability margins characterize the reserve of the loop with respect to gain and phase variations, while the bandwidth defines the dynamic range over which the closed-loop axis response can follow reference commands and attenuate low-frequency disturbances. This frequency-domain assessment verifies that the intended main-axis control architecture provides sufficient robustness before being applied in the time-domain pointing simulations.

\begin{figure}[ht]
    \vspace{.4cm}
    \centering
    \begin{subfigure}[t]{0.48\linewidth}
        \centering
        \includegraphics[width=\linewidth]{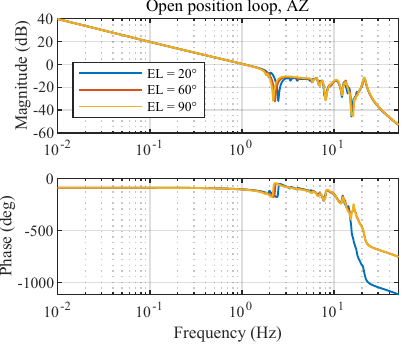}
        \caption{Azimuth axis}
        \label{fig:open_loop_bode_az}
    \end{subfigure}
    \hfill
    \begin{subfigure}[t]{0.48\linewidth}
        \centering
        \includegraphics[width=\linewidth]{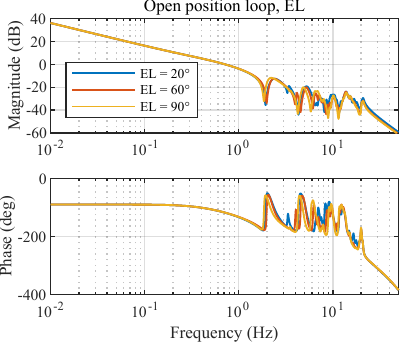}
        \caption{Elevation axis}
        \label{fig:open_loop_bode_el}
    \end{subfigure}
    \vspace{.2cm}
    \caption{Open-loop Bode plots of the azimuth and elevation position-control loops used for tuning the cascaded main-axis controller. The plots are evaluated on the flexible telescope mount model and provide the basis for the gain and phase margins summarized in Table~\ref{tab:main_axis_control_characteristics}.}
    \label{fig:open_loop_bode}
\end{figure}

Representative open-loop Bode plots for the azimuth and elevation position-control loops are shown in Fig.~\ref{fig:open_loop_bode}. The plots illustrate the tuned loop dynamics used in the end-to-end simulations and provide the basis for the stability margins summarized in Table~\ref{tab:main_axis_control_characteristics}.

\begin{figure}[ht]
    \vspace{.4cm}
    \centering
    \begin{subfigure}[t]{0.48\linewidth}
        \centering
        \includegraphics[width=\linewidth]{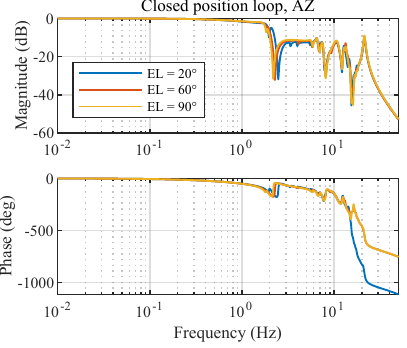}
        \caption{Azimuth axis}
        \label{fig:closed_loop_bode_az}
    \end{subfigure}
    \hfill
    \begin{subfigure}[t]{0.48\linewidth}
        \centering
        \includegraphics[width=\linewidth]{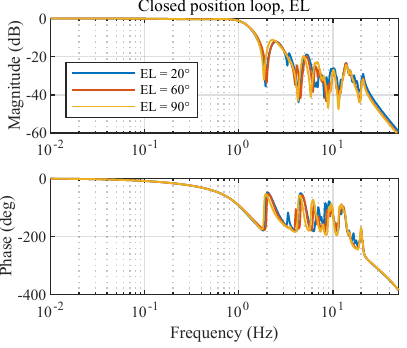}
        \caption{Elevation axis}
        \label{fig:closed_loop_bode_el}
    \end{subfigure}
    \vspace{.2cm}
    \caption{Closed-loop Bode plots of the azimuth and elevation position-control loops. The curves show the reference-to-measured-position response of the tuned main-axis controllers for the investigated elevation configurations.}
    \label{fig:closed_loop_bode}
\end{figure}

The resulting closed-loop reference responses are shown in Fig.~\ref{fig:closed_loop_bode}. These responses provide the effective position-loop bandwidths used to characterize the tracking dynamics of the tuned main-axis controllers.

\begin{table}[ht]
\centering
\caption{Control characteristics of the tuned main-axis position-control loops used in the end-to-end dynamic pointing simulations.}
\label{tab:main_axis_control_characteristics}
\vspace{.2cm}
\renewcommand{\arraystretch}{1.15}
\begin{tabular}{lccc ccc}
\toprule
& \multicolumn{3}{c}{Azimuth axis} & \multicolumn{3}{c}{Elevation axis} \\
\cmidrule(lr){2-4} \cmidrule(lr){5-7}
Metric 
& EL=$20^\circ$ & EL=$60^\circ$ & EL=$90^\circ$
& EL=$20^\circ$ & EL=$60^\circ$ & EL=$90^\circ$ \\
\midrule
Phase margin
& $74.3^\circ$ & $73.0^\circ$ & $72.3^\circ$
& $61.9^\circ$ & $63.0^\circ$ & $63.0^\circ$ \\

Gain margin
& $12.4\,\mathrm{dB}$ & $12.5\,\mathrm{dB}$ & $11.5\,\mathrm{dB}$
& $25.1\,\mathrm{dB}$ & $25.7\,\mathrm{dB}$ & $25.3\,\mathrm{dB}$ \\

Bandwidth
& $1.6\,\mathrm{Hz}$ & $1.6\,\mathrm{Hz}$ & $1.5\,\mathrm{Hz}$
& $1.18\,\mathrm{Hz}$ & $1.18\,\mathrm{Hz}$ & $1.18\,\mathrm{Hz}$ \\
\bottomrule
\end{tabular}
\end{table}

The resulting controller tuning provides stable closed-loop behaviour over the investigated elevation range. The azimuth loop achieves a slightly higher position bandwidth than the elevation loop, while both axes retain substantial phase and gain margins. These bandwidth values define the frequency range in which the main-axis control system can actively contribute to disturbance rejection and trajectory tracking, and are therefore important for the interpretation of both wind-induced and scan-induced dynamic pointing errors.

\section{Wind disturbance modelling}
\label{sec:wind_modelling}

Wind loading is a critical external disturbance for the dynamic pointing performance of large radio telescopes. For AtLAST, the relevant quantity is not only the mean wind speed or a quasi-static design load, but the time-dependent wind excitation in the frequency range in which the flexible telescope structure and the main-axis control system are dynamically responsive. The wind model used in this work is therefore derived from measured site-specific wind time series and applied as dynamic input to the end-to-end telescope simulation.

A key aspect of the present analysis is that the wind excitation is derived from dedicated, high-rate measurements obtained at the AtLAST candidate sites before telescope construction. Such site-specific, time-resolved wind data are rarely available at this stage of a large-telescope design and provide an unusually direct link between site characterization and dynamic performance verification. As part of the AtLAST site-characterization campaign, meteorological towers equipped with sonic anemometers were deployed on the Chajnantor Plateau to characterize the local wind field and its relevance for telescope design and operation~\cite{atlast_site_selection_report,pizarro_inprep}. The measured time series provide access not only to wind-speed statistics, but also to the spectral content of the atmospheric excitation, enabling the dynamic pointing performance to be evaluated against realistic site-dependent wind input rather than generic turbulence assumptions.

This is particularly important for dynamic pointing. Although AtLAST is a very massive structure, with a total telescope mass on the order of several thousand tonnes, the pointing requirement is governed by small angular deviations of the optical axis. Consequently, even small wind-induced structural deformations and rotations can contribute to measurable pointing jitter. The resulting dynamic pointing error depends on the coupling between the wind-load spectrum, the flexible structural modes of the mount, and the finite bandwidth of the main-axis control system. Preserving the measured temporal structure of the wind excitation is therefore essential for a meaningful end-to-end verification of the telescope design.

\subsection{Wind-data preparation}
\label{subsec:wind_data_preparation}

The wind data used for the dynamic pointing simulations are derived from the high-rate site measurements described in the AtLAST site-selection documentation and in the ongoing wind-data analysis by Pizarro et al.~\cite{atlast_site_selection_report,pizarro_inprep}. The measurements provide time series of wind speed and wind direction at the candidate sites. For the dynamic simulations, the high-rate wind data are used in the time domain in order to retain the spectral content relevant for structural excitation and control-loop interaction.

For the end-to-end simulations, the measured wind time series are divided into blocks of $1000\,\mathrm{s}$. This duration provides a sufficiently long realization of the atmospheric excitation for estimating dynamic pointing statistics, while keeping the simulation campaign computationally tractable. The block-wise representation also enables a systematic classification of the measured wind conditions according to their RMS wind speed, while preserving the temporal structure relevant for dynamic structural excitation.

For each accepted block, the root-mean-square wind speed is computed and used to assign the block to a discrete wind-speed class. The data set is grouped into wind-speed levels with approximately $1\,\mathrm{m\,s^{-1}}$ increments, spanning low-wind conditions up to wind speeds exceeding $19\,\mathrm{m\,s^{-1}}$. Representative blocks from each wind-speed class are selected as input to the end-to-end simulation. This approach preserves realistic temporal wind fluctuations while enabling a systematic evaluation of pointing performance as a function of wind speed.

\subsection{Wind-load generation for the end-to-end simulation}
\label{subsec:wind_load_generation}

For each selected wind block, the measured horizontal wind speed and wind direction are used to generate time-dependent aerodynamic loads for the telescope model. The wind velocity is converted into dynamic pressure and combined with the aerodynamic load coefficients available from the structural model. This yields time-dependent forces and moments acting on the exposed telescope components.

The wind direction is treated consistently with the aerodynamic load model, allowing different wind angles of attack to be represented in the simulations. The resulting load time series are applied to the flexible state-space model of the telescope mount together with the actuator torques generated by the main-axis control system. In this way, the measured site wind conditions enter the end-to-end simulation as dynamic external loads, while the structural and control response is computed by the coupled telescope model.

\subsection{Tracking simulations under wind loading}
\label{subsec:wind_tracking_simulations}

Wind-induced dynamic pointing is evaluated under closed-loop tracking conditions. In this simulation case, the commanded azimuth and elevation positions are held constant, while the measured wind time series are applied as external dynamic loads to the flexible telescope model. The main-axis control system therefore acts to maintain the commanded telescope orientation in the presence of realistic site-dependent wind excitation. No scanning motion is applied, so that the resulting pointing error can be attributed to wind-induced excitation and the corresponding closed-loop structural response.

The simulations are performed for telescope elevations of $20^\circ$, $60^\circ$, and $90^\circ$. For each elevation, multiple wind-speed classes and wind angles of attack are evaluated. The pointing error is computed in cross-elevation and elevation coordinates and expressed as RMS dynamic pointing error after removal of the static component according to Eq.~\eqref{eq:dynamic_pointing_error}. In addition to the total pointing error, the position-control error of the main axes is evaluated as a diagnostic quantity. This quantity represents the part of the response associated with finite tracking accuracy of the controlled axes and is used to distinguish control-related effects from residual pointing errors caused by flexible structural deformation.

\begin{figure}[ht]
    \centering
    \includegraphics[width=0.85\linewidth]{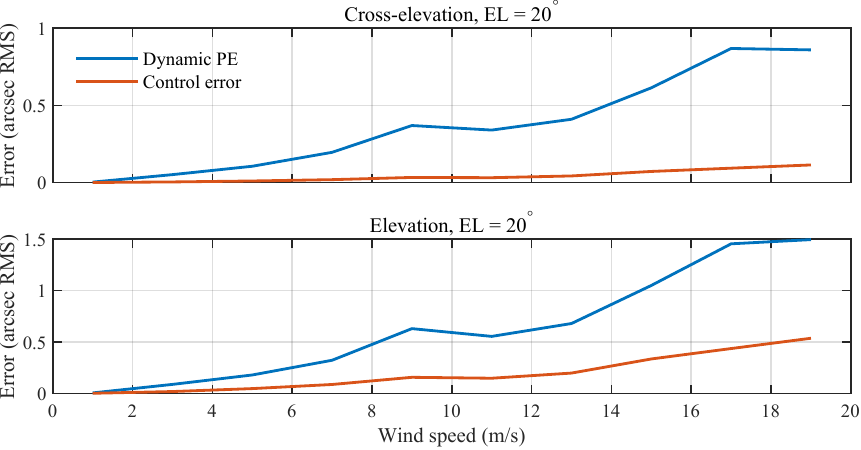}
    \vspace{.1cm}
    \caption{Mean wind-induced dynamic pointing error and position-control contribution as a function of wind speed for EL~$=20^\circ$, averaged over the investigated wind angles of attack. The two panels show the cross-elevation and elevation components separately.}
    \label{fig:pe_wind_el20}
\end{figure}

Figure~\ref{fig:pe_wind_el20} shows the mean wind-induced dynamic pointing error (PE) and the corresponding position-control contribution as a function of wind speed for EL~$=20^\circ$. This elevation is the most critical of the investigated tracking configurations. The total dynamic pointing error increases with wind speed in both cross-elevation and elevation. The position-control contribution follows the same general trend but represents only a limited part of the total response. The remaining dynamic pointing error is therefore attributed primarily to the residual flexible response of the telescope structure under wind loading.

Table~\ref{tab:wind_tracking_summary_15ms} summarizes the elevation dependence for a representative strong-wind case of $15\,\mathrm{m\,s^{-1}}$. This wind level is used in the tracking-stability budget comparison and therefore provides a direct link between the wind-induced response and the design verification. The values are averaged over the investigated wind angles of attack. The largest dynamic pointing error occurs at EL~$=20^\circ$, reaching $0.61\,\mathrm{arcsec}$ RMS in cross-elevation and $1.05\,\mathrm{arcsec}$ RMS in elevation. The corresponding values decrease to $0.39\,\mathrm{arcsec}$ RMS and $0.68\,\mathrm{arcsec}$ RMS at EL~$=60^\circ$, and to $0.30\,\mathrm{arcsec}$ RMS and $0.52\,\mathrm{arcsec}$ RMS at EL~$=90^\circ$.

\begin{table}[ht]
\centering
\caption{Mean wind-induced dynamic pointing error and position-control contribution at $15\,\mathrm{m\,s^{-1}}$, averaged over wind angle of attack.}
\label{tab:wind_tracking_summary_15ms}
\vspace{.2cm}
\renewcommand{\arraystretch}{1.15}
\begin{tabular}{lcccc}
\toprule
& \multicolumn{2}{c}{Dynamic pointing error} & \multicolumn{2}{c}{Position-control contribution} \\
\cmidrule(lr){2-3} \cmidrule(lr){4-5}
Telescope elevation & XEL & EL & XEL & EL \\
& arcsec RMS & arcsec RMS & arcsec RMS & arcsec RMS \\
\midrule
$20^\circ$ & 0.61 & 1.05 & 0.07 & 0.33 \\
$60^\circ$ & 0.39 & 0.68 & 0.02 & 0.32 \\
$90^\circ$ & 0.30 & 0.52 & 0.00 & 0.30 \\
\bottomrule
\end{tabular}
\end{table}
\FloatBarrier

The comparison in Table~\ref{tab:wind_tracking_summary_15ms} shows that the control-related contribution is small in cross-elevation and more pronounced in elevation. This behaviour is consistent with the different closed-loop dynamics of the azimuth and elevation axes discussed in Sec.~\ref{sec:control_design}. However, the wind-induced pointing response is not governed by axis tracking alone. The diagnostic comparison between total pointing error and position-control contribution indicates that flexible structural deformation is the dominant mechanism in the tracking case under wind loading.

The wind-induced dynamic pointing error is finally compared with the allocated tracking-stability budget in Table~\ref{tab:wind_budget_comparison}. The comparison is performed for EL~$=20^\circ$, which is the most critical of the investigated elevations, and uses the absolute dynamic pointing error. The end-to-end simulation remains below the allocation for all listed wind speeds. At the strongest budgeted wind condition of $15\,\mathrm{m\,s^{-1}}$, the simulated dynamic pointing error is $1.21\,\mathrm{arcsec}$ RMS compared with an allocation of $5.5\,\mathrm{arcsec}$.

Together, the budget comparison and response decomposition verify that the current AtLAST design satisfies the wind-induced dynamic pointing allocation under realistic site wind excitation. The remaining wind-induced pointing error is mainly associated with flexible structural response, while the position-control contribution is not the limiting factor.

\begin{table}[ht]
\centering
\vspace{.2cm}
\caption{Comparison of wind-induced dynamic pointing error with the allocated tracking-stability budget for EL~$=20^\circ$.}
\label{tab:wind_budget_comparison}
\renewcommand{\arraystretch}{1.15}
\begin{tabular}{lccc}
\toprule
Wind speed & Allocated budget & E2E simulation & Margin factor \\
& arcsec & arcsec RMS & -- \\
\midrule
$3.5\,\mathrm{m\,s^{-1}}$ & 0.30 & 0.25 & 1.2 \\
$9\,\mathrm{m\,s^{-1}}$   & 2.00 & 0.73 & 2.7 \\
$15\,\mathrm{m\,s^{-1}}$ & 5.50 & 1.21 & 4.5 \\
\bottomrule
\end{tabular}
\end{table}
\FloatBarrier

\section{Scan-induced dynamic pointing during fast mapping}
\label{sec:scan_induced_pointing}

Fast mapping introduces self-induced dynamic pointing errors because the commanded scan trajectory generates actuator torques that interact with the flexible mount dynamics. In contrast to tracking under wind loading, the excitation is therefore generated by the closed-loop telescope motion itself. To isolate this effect from wind loading, the scan simulations are performed without external wind loads. This separation follows the modelling strategy introduced in Sec.~\ref{subsec:simulation_cases}: wind-induced and scan-induced effects are evaluated independently in order to identify their relative magnitudes and physical origins. While real observations may combine fast scanning with simultaneous wind loading, the isolated scan case provides a clear assessment of whether the fast-scan response is primarily limited by main-axis tracking bandwidth or by residual structural flexibility.

\subsection{Lissajous-daisy scan definition}
\label{subsec:lissajous_daisy_scan}

\begin{figure}[ht]
    \centering
    \includegraphics[width=0.7\linewidth]{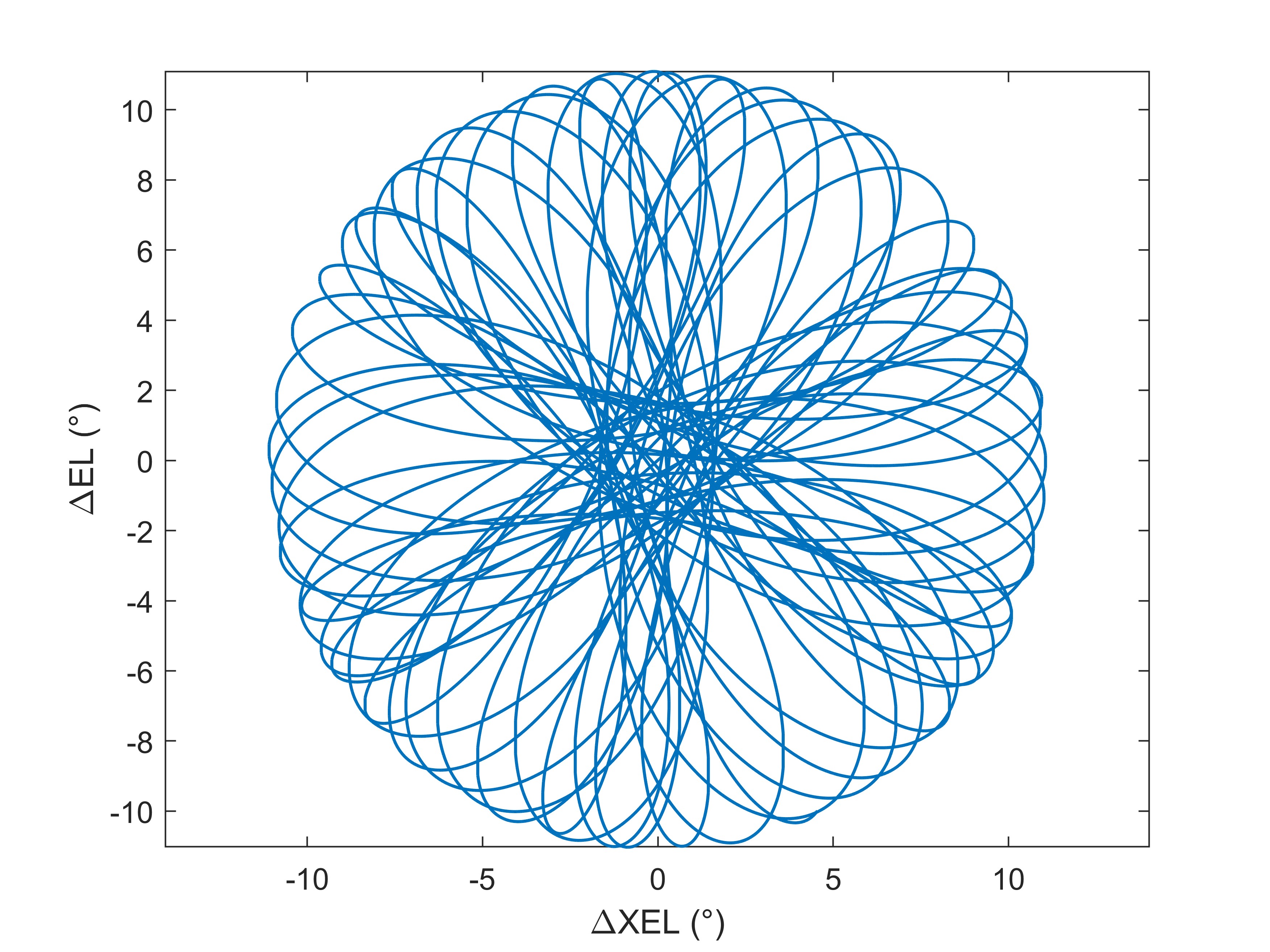}
    \vspace{.1cm}
    \caption{Representative Lissajous-daisy scan trajectory used for the scan-induced dynamic pointing analysis. The trajectory is shown in cross-elevation and elevation coordinates and represents a demanding fast-mapping case close to the kinematic limits of the mount.}
    \label{fig:lissajous_daisy_scan}
\end{figure}

The scan-induced pointing analysis uses the Lissajous-daisy trajectory shown in Fig.~\ref{fig:lissajous_daisy_scan}. This trajectory represents a demanding fast-mapping case and is constrained by a maximum angular velocity of $3\,\mathrm{deg\,s^{-1}}$ and a maximum angular acceleration of $1\,\mathrm{deg\,s^{-2}}$. The scan is therefore used to assess the self-induced dynamic pointing response of the closed-loop telescope model under high excitation.

The scan is evaluated for telescope elevations of $20^\circ$, $30^\circ$, and $45^\circ$. Higher elevation angles are not considered for this scan pattern because the corresponding azimuth motion would exceed the applicable velocity limits. The resulting pointing errors are evaluated in cross-elevation and elevation coordinates using the same RMS definition as in Eq.~\eqref{eq:dynamic_pointing_error}.

The Lissajous-daisy scan should be interpreted as a deliberately demanding scan case. The resulting self-induced pointing error is therefore not representative of all possible observing patterns, but provides a conservative assessment of the telescope response under a fast, high-excitation trajectory.

\subsection{Scan-induced pointing-error decomposition}
\label{subsec:scan_pointing_decomposition}

For the scan simulations, the total dynamic pointing error is evaluated together with the position-control contribution and the residual contribution. The position-control contribution characterizes the finite tracking accuracy of the controlled azimuth and elevation axes while following the commanded scan trajectory. The residual contribution is obtained from the time-domain pointing-error signals and is dominated by flexible structural deformation that is not represented by the controlled axis motion alone. As discussed in Sec.~\ref{subsec:pointing_error_definitions}, this decomposition is performed before RMS evaluation. Therefore, the residual RMS value is not generally equal to a scalar difference between the RMS values of total pointing error and position-control contribution.

\begin{figure}[ht]
    \centering
    \includegraphics[width=0.9\linewidth]{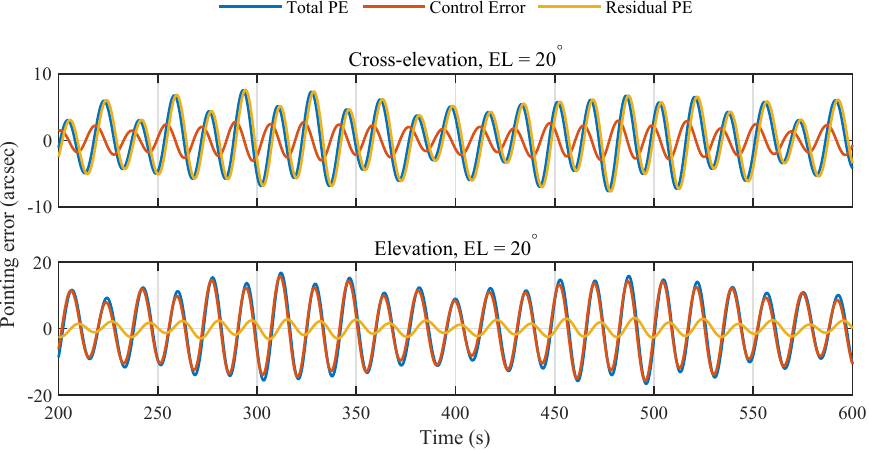}
    \vspace{.1cm}
    \caption{Time-domain pointing-error decomposition during the Lissajous-daisy scan for EL~$=20^\circ$. The total pointing error, position-control contribution, and residual contribution are shown separately for cross-elevation and elevation.}
    \label{fig:lissajous_scan_pointing_error_el20}
\end{figure}

Figure~\ref{fig:lissajous_scan_pointing_error_el20} shows the time-domain pointing-error decomposition for the representative scan case at EL~$=20^\circ$. In elevation, the total pointing error closely follows the position-control contribution, indicating that the scan-induced elevation response is mainly associated with finite trajectory-tracking accuracy of the main axes. In cross-elevation, the residual contribution is more pronounced and closely follows the total pointing-error response over large parts of the scan. This shows that scan-induced excitation of flexible structural modes remains relevant, particularly in cross-elevation.

The corresponding RMS values are summarized in Table~\ref{tab:lissajous_scan_results}. In elevation, the total pointing error is largely associated with the position-control contribution. At EL~$=20^\circ$, for example, the total elevation pointing error is $9.26\,\mathrm{arcsec}$ RMS, while the corresponding position-control contribution is $8.58\,\mathrm{arcsec}$ RMS. Similar values are obtained at EL~$=30^\circ$ and EL~$=45^\circ$, confirming that the elevation response during this fast scan is primarily governed by the ability of the main-axis control system to follow the commanded trajectory.

\begin{table}[ht]
\centering
\caption{Pointing-error decomposition for the Lissajous-daisy scan. Values are given in arcsec RMS.}
\label{tab:lissajous_scan_results}
\vspace{.2cm}
\renewcommand{\arraystretch}{1.15}
\begin{tabular}{lcccccc}
\toprule
& \multicolumn{2}{c}{EL~$=20^\circ$} & \multicolumn{2}{c}{EL~$=30^\circ$} & \multicolumn{2}{c}{EL~$=45^\circ$} \\
\cmidrule(lr){2-3} \cmidrule(lr){4-5} \cmidrule(lr){6-7}
Quantity & XEL & EL & XEL & EL & XEL & EL \\
\midrule
Total pointing error & 3.97 & 9.26 & 3.69 & 9.18 & 3.53 & 9.36 \\
Position-control contribution & 1.66 & 8.58 & 1.69 & 8.58 & 1.72 & 8.58 \\
Residual contribution & 4.01 & 1.63 & 3.68 & 0.98 & 3.45 & 1.10 \\
\bottomrule
\end{tabular}
\end{table}
\FloatBarrier

In cross-elevation, the residual contribution is more pronounced. At EL~$=20^\circ$, the total cross-elevation pointing error is $3.97\,\mathrm{arcsec}$ RMS, while the residual contribution is $4.01\,\mathrm{arcsec}$ RMS. As noted above, this does not imply that the residual contribution is a scalar component larger than the total response. Rather, it reflects the fact that the decomposition is performed on the time-domain signals before RMS evaluation. The result indicates that flexible structural deformation contributes significantly to the scan-induced cross-elevation pointing response.

This distinction is relevant for future correction strategies. The position-control contribution is derived from the known command trajectory and the measured main-axis response. It is therefore observable and can be used in correction strategies, either through feedforward or trajectory pre-shaping for predictable components, or through map-domain post-processing when time-stamped tracking errors are available. The residual contribution is different in nature, because it is associated with flexible structural deformation and cannot be inferred from the main-axis tracking error alone.

A dedicated pointing-error allocation for self-induced fast-scan jitter is still to be consolidated in the next design phase. The present analysis provides the quantitative basis for this by identifying the magnitude and physical origin of the scan-induced pointing errors. The Lissajous-daisy case is therefore interpreted as a design-driving scenario for future scan-mode-specific requirements and control optimization, with feedforward, trajectory shaping, scan-specific correction, map-domain post-processing, and control-loop retuning identified as relevant mitigation paths.

\section{Discussion}
\label{sec:discussion}

The end-to-end simulations provide two main outcomes. First, they verify the current AtLAST design with respect to the allocated dynamic pointing performance under wind loading. Second, they separate the dominant mechanisms that contribute to self-induced pointing errors during fast scanning. This distinction is important because tracking under wind loading and fast mapping are governed by different combinations of structural response, control bandwidth, and trajectory-dependent excitation.

For tracking under wind loading, the simulated dynamic pointing error remains within the allocated tracking-stability budget for all budgeted wind-speed levels. Since the wind input is derived from high-rate (20~Hz) site measurements, this provides a direct design-verification link between the measured environmental conditions at the candidate sites and the predicted dynamic pointing performance. The response decomposition shows that the remaining wind-induced pointing error is mainly associated with flexible structural deformation, while the position-control contribution is not the limiting factor. Since this residual response remains within the allocated budget, the results support the adequacy of the current AtLAST structural concept and main-axis control approach for dynamic pointing stability under wind loading.

The scan-induced case should be interpreted differently. During the Lissajous-daisy scan, finite main-axis bandwidth leads to trajectory-following errors, while the actuator torques and accelerations associated with the commanded motion can independently excite flexible structural modes. The elevation response is mainly associated with the position-control contribution, consistent with the lower position-loop bandwidth of the elevation axis. In cross-elevation, the residual flexible contribution remains significant. Fast mapping is therefore a coupled control--structure problem, but with a stronger role of trajectory tracking and control bandwidth than in the wind-loaded tracking case.

This interpretation has direct consequences for the next design steps. The wind results constitute a budget-compliance result for the current design under realistic measured wind excitation. The scan results, by contrast, provide the quantitative basis for defining future scan-mode-specific pointing allocations. Since the position-control contribution is observable from the commanded and measured main-axis motion, it can be addressed through scan-specific correction and/or map-domain post-processing of time-stamped tracking errors.

\section{Conclusion}

An end-to-end dynamic pointing analysis has been performed for the current AtLAST telescope concept under wind loading and fast-scan conditions. The analysis combines a flexible structural model, the intended cascaded main-axis control architecture, pointing-error evaluation, and high-rate measured wind excitation from the AtLAST candidate sites.

The tracking simulations under wind loading show that the wind-induced dynamic pointing error remains within the allocated tracking-stability budget for all investigated budgeted wind conditions. This confirms the adequacy of the current AtLAST structural concept and main-axis control approach with respect to dynamic pointing stability under wind loading. The use of measured site wind data provides a direct link between site characterization and design verification.

The fast Lissajous-daisy scan simulations reveal a different error composition. In elevation, the scan-induced pointing error is mainly associated with finite main-axis control bandwidth and trajectory-following accuracy. This contribution is observable through the commanded and measured main-axis motion and can therefore be addressed by control-oriented correction strategies or map-making procedures. In cross-elevation, residual flexible structural response remains significant and requires structural-dynamic modelling or active compensation concepts.

The scan case therefore identifies fast mapping as a coupled control--structure problem. Since dedicated pointing-error allocations for self-induced fast-scan jitter are still to be consolidated, the presented analysis provides a quantitative basis for future scan-mode-specific requirements and optimization strategies, including feedforward, trajectory pre-shaping, map-domain post-processing of time-stamped tracking errors, scan-specific correction, and control-loop retuning.

Overall, the results support the feasibility of the current AtLAST design for high-speed submillimeter mapping from a dynamic pointing perspective. The proposed end-to-end simulation framework verifies compliance with wind-induced pointing allocations and identifies the dominant mechanisms and optimization paths for future fast-scan operation.

\acknowledgments
This project has received funding from the European Union's Horizon Europe research and innovation programme under grant agreement No. 101188037 (AtLAST2). Views and opinions expressed are however those of the author(s) only and do not necessarily reflect those of the European Union or European Research Executive Agency. Neither the European Union nor the European Research Executive Agency can be held responsible for them.

TM acknowledges support from the Agencia Estatal de Investigaci\'on (AEI) and the Ministerio de Ciencia, Innovaci\'on y Universidades (MICIU) Grant ATRAE2024-154740 funded by MICIU/AEI//10.13039/501100011033. TM is also partly supported by the Spanish program Unidad de Excelencia María de Maeztu CEX2020-001058-M, financed by MCIN/AEI//10.13039/501100011033, and by the MaX-CSIC Excellence Award MaX4-SOMMA-ICE.

The authors acknowledge the use of AI-assisted language tools for grammar checking, sentence rephrasing, and improving the readability of selected parts of the manuscript. All technical content, data analysis, interpretation of results, figures, and conclusions were developed, reviewed, and approved by the authors.

\bibliographystyle{spiebib}
\bibliography{report}

\end{document}